\documentclass[useAMS,usenatbib]{mn2e}

\usepackage{graphicx}

\title[Low angular momentum flow model for Sgr A*]
 {Low angular momentum flow model for Sgr A*}
\author[T. Okuda and D. Molteni]{T. Okuda$^{1}$
\thanks{E-mail:bbnbh669@ybb.ne.jp}, D. Molteni$^{2}$ \\
$^{1}$ Nishi-Asahioka-Cho 3-15-1, Hakodate 042-0915, Hokkaido, Japan\\
$^{2}$Dipartimento di Fisica, Universit\`{a} di Palermo,
Viale delle Scienze,  Palermo, 90128, Italy}

\begin{document}

\date{Accepted }

\pagerange{\pageref{0}--\pageref{0}} \pubyear{2012}

\maketitle

\label{firstpage}

\begin{abstract}
We examine the low angular momentum flow model for Sgr A*
using two-dimensional  hydrodynamical calculations based on 
the parameters of the specific angular momentum and total energy 
estimated in the recent analysis of stellar wind of nearby stars 
around Sgr A*.
The accretion flow with the plausible parameters is non-stationary 
and an irregularly oscillating shock is formed in the inner region 
of a few tens to a hundred and sixty Schwarzschild radii.
Due to the oscillating shock, the luminosity and  the mass-outflow
rate are modulated by several per cent to a factor of 5 and a factor of 2-7, 
 respectively, on time-scales of an hour to 
ten days. The flows are highly advected and the radiative efficiency of the
 accreting matter into radiation is very low, $\sim 10^{-5}$--$10^{-3}$, and
 the input accretion rate $\dot M_{\rm input} = 4.0\times
 10^{-6}\;{\rm M_{\odot}}$ yr$^{-1}$ results in the observed 
luminosities $\sim 10^{36}$ erg s$^{-1}$ of Sgr A* if a two-temperature model
 and the synchrotron emission are taken into account.
The mass-outflow rate of the gas originating in the post-shock
region increases with the increasing input specific angular momentum 
and ranges from a few to 99 per cent of the input accreting matter, 
depending on the input angular momentum.
The oscillating  shock is necessarily triggered if the specific angular
momentum and the specific energy belong to or are located just nearby 
in the range of parameters responsible for a stationary shock in rotating inviscid and  adiabatic accretion flow. 
The time variability may be relevant to the flare activity of Sgr A*.

\end{abstract}

\begin{keywords}
accretion, accretion discs -- black hole physics -- hydrodynamics -- 
radiation mechanism: thermal -- shock waves -- Galaxy: centre.

\end{keywords}

\section{Introduction}
Disc accretion is an essential process for such phenomena as energetic
X-ray sources, active galactic nuclei and protostars.
Since the early works by \citet{b25} and \citet{b26}, a great number 
of papers have been devoted to studies of the disc accretion onto 
gravitating objects. 
When the accretion rate is not too high, the accretion disc luminosity
is directly in proportion to the accretion rate and can be successfully
described by the standard thin disc model (S-S model).
Sgr A* in our Galactic centre has been extensively studied in the 
category of accretion processes because it is a supermassive black hole
 candidate in our Galaxy and has unique observational features incompatible
 with the S-S model.
One of the features of Sgr A* is that the observed luminosity 
is five orders of magnitude lower than that predicted by the S-S model.
Moreover, the spectrum of Sgr A* differs from the multi-temperature black body
spectrum obtained from the S-S model.

Since the observational features of Sgr A* can not be explained by the S-S
model, two types of theoretical models, such as the  spherical Bondi accretion
model without any net angular momentum \citep{b1} and the advection-dominated 
accretion flow model (ADAF) with high angular momentum \citep{b19,b20}, 
have been proposed (Narayan \& McClintock 2008, Yuan 2011 for review). 
Both the Bondi model and the ADAF model result in highly advected flows and
 the radiative efficiency is so low as to be compatible with the observations.
The spherically symmetric Bondi model gave the first detailed model of 
Sgr A* \citep{b10,b11}.
This model applies when the circularization radius is smaller than the 
innermost stable circular orbit. The Bondi model is frequently used to 
estimate the expected accretion rate near the Bondi radius far from the 
gravitating object.
However, the simple Bondi model can not be extended close to the black hole to
satisfy the details of the activity of Sgr A*. 
In contrast with the Bondi model, the ADAF model applies  when the angular 
momentum at the outer boundary is roughly Keplerian, and a large fraction
of the viscously generated heat is advected with the accreting gas and only 
a small fraction of the energy is radiated.
 Unlike the cold standard disc, this model results in a high temperature,
rapid accretion, sub-Keplerian disc and inefficient radiation in the inner
 region of the disc.
The ADAF models were shown to be generally successful and more advanced models
\citep{b31,b32}, taking into account the parametric description of the outflow
 and jet, explain well the observations.
The important key to these models for Sgr A* is the amount of angular 
momentum in the accretion flow.
However, at present, we have no clear evidence for the angular momentum from
observations.

The low angular momentum model belongs to an intermediate case between the 
Bondi model and the ADAF model and was applied to Sgr A* \citep {b17,b3}. 
Assuming that the Wolf-Rayet star ${\rm IRS}$ 13 $ \rm E3$ is the dominant
source of the matter accreting onto Sgr A* and assuming the wind
temperature $T_{\rm wind}$ = 1.0 or 0.5 keV, they estimated the net 
angular momentum  $\lambda$ of 1.68--2.16 and the Bernoulli constant 
$\varepsilon$ of $3.97 \times 10^{-6}$--$1.98 \times 10^{-6}$, 
where the mass $M$, the speed of light $c$ and the Schwarzschild radius 
$R_{\rm g}=2GM/c^2$ are used as the units of mass, velocity and distance. 
With these flow parameters for $\lambda$ and $\varepsilon$,
they showed analytically that there is no continuous flow solution which 
attains to the event horizon, and the resulting flow would be non-stationary, 
but that, for the case of the angular momentum 
$\lambda$= 1.55 lower than the best estimates for Sgr A*, there exists a 
standard stationary shock solution. 
Subsequently, they speculated that, for the non-stationary flow, 
the inflowing matter accumulates and forms a 
ring or torus and that the ring in the inner region is unstable and may 
explain the variability of Sgr A*. Motivated by their suggestion and results, 
we examine the low angular momentum flow model for Sgr A* using 2D  
hydrodynamical calculations and discuss the results on the activity of Sgr A*.

\section{Modelling a low angular momentum flow}
We consider a supermassive black hole with mass $M= 4\times 
10^6 M_{\odot}$ for Sgr A*.
For the accretion flow around Sgr A*,  we use the flow parameters 
$\lambda$ and $\varepsilon$ estimated by \citet{b17} and the mass accretion 
rate $\dot M=4.0 \times 10^{-6}M_{\odot}$ yr$^{-1}$ in all models.
1D flow solution of a thin, rotating, inviscid and adiabatic accretion flow
 near the equatorial plane generally depends on the flow model especially on
 its thickness, for example, the vertical equilibrium model or the constant
 height model.
Assuming the vertical equilibrium model here, we solve the Bernoulli equation 
with the constant angular momentum $\lambda$, find the outer and inner sonic
 points and get the Mach number versus radius relation, the sound speed 
$v_{\rm s}$, the thickness $h$ of the accretion flow, the radial velocity $v$
 and the Temperature $T$ at a given radius $r$ \citep{b2}.
It should be noticed that the configuration of the Mach number versus radius 
does not coincide completely with that in \citet{b17} because of 
the different flow model used. 
The Mach number versus radius relations for $\lambda = 1.68$ and 2.16 are 
similar to theirs, that is, the flow which passes through the 
outer sonic point never continue to the event horizon.
The flow for $\lambda= 1.55$ also shows a flow topology similar
to that for $\lambda=1.68$, unlike the case of \citet{b17}.
 The first purpose in this study is to examine how the flow with these specific
angular momenta actually behaves.
However, in addition to the above parameters, we examined another case
of $\lambda = 1.35$, for which the standing shock is predicted to exist in the
inner region.

Figs ~1 and 2 show the Mach number versus radius in 1D adiabatic flow 
obtained for the cases of $\lambda$ = 1.68 and $\varepsilon$ = $3.97 
\times 10^{-6}$ (model A) and $\lambda$ = 1.35 and $\varepsilon$ = $1.98 
\times 10^{-6}$ (model D), respectively.
In Fig.~1, the particle which passes through the outer sonic point A  falls 
down supersonically inward but never attains the event horizon since it
makes a closed loop of the Mach number curve.
On the other hand, in Fig.~2, the particle falls supersonically along the 
solid line AB, jumps at the shock position of $R_{\rm s} \sim 20 R_{\rm g}$ 
from the point B to C in a subsonic state and tends supersonically toward 
the event horizon along the dashed line CD, where the Mach number curve 
(dashed line) which passes through the inner sonic point intersects at the 
point C with the other Mach number curve AC (dotted line) which passes through 
the outer sonic point and satisfies the Rankine-Hugoniot relation between 
the solid and the dashed lines. In Fig. 2, there exist two shock positions 
($R_{\rm s} \sim$ 17 and 19). It is known that usually the inner 
shock of the two shocks is unstable but the outer one is stable \citep{b18}.

\begin{table*}
\centering
\caption{Model parameters  for the accreting matter onto Sgr A*, where
 in models E and F a two-temperature model is adopted and $\delta$ is the 
 initial ratio $T_{\rm e}/T_{\rm i}$ of electron temperature $T_{\rm e}$ to
 ion temperature $T_{\rm i}$.}
\begin{tabular}{@{}ccccccccc} \hline \hline
model&$\lambda$ & $\varepsilon$& $\gamma$& $\dot M$ $(M_{\odot}$ 
 ${\rm yr}^{-1})$ & $\delta$                                    \\\hline
 $A$    & 1.68  & 3.97 $\times 10^{-6}$ & 1.6& 4.0 $\times 10^{-6}$ & -- \\
 $B$   & 2.16  & 1.98 $\times 10^{-6}$ & 1.6 &4.0 $\times 10^{-6} $ & -- \\
 $C$   & 1.55  & 1.98 $\times 10^{-6}$ & 1.6 & 4.0 $\times 10^{-6}$ & -- \\
 $D$    & 1.35  & 1.98 $\times 10^{-6}$ &1.6  &4.0 $\times 10^{-6}$ & -- \\
 $E$    & 1.68  & 3.97 $\times 10^{-6}$ &1.6  &4.0 $\times 10^{-6}$ & 3/7\\
 $F$    & 1.35  & 1.98 $\times 10^{-6}$ &1.6  &4.0 $\times 10^{-6}$ & 1/9\\
\hline
\end{tabular}
\end{table*}

Thus far, the model parameters are listed in Table 1, 
where $\gamma$ is the adiabatic index.
First we examine models A--D using 2D radiation hydrodynamical calculations
which include only the free-free transitions in the cooling and heating rate.
It is well known that the existence of the sub-millimetre bump in Sgr A* 
spectra is certainly caused by the synchrotron emission. Secondly, following
the results of models A and D, we examine another models of E and F which
 take account of the synchrotron cooling in the energy equation under an
 assumption that the magnetic energy density is 10 per cent of the thermal
 energy density.
 In models E and F, we adopt a two-temperature model where the ratio 
$T_{\rm e}/T_{\rm i}$ of the electron temperature $T_{\rm e}$ to the
ion temperature $T_{\rm i}$ is initially assumed to be  a constant 
$\delta(\leq 1)$ and neglect the radiation transport because the flow is 
sufficiently optically thin and the flux-limitted diffusion approximation of
 radiation is less accurate for so optically thin medium.

\begin{figure}
\begin{minipage}{0.5\linewidth}
\includegraphics[width=7cm, height=6cm]{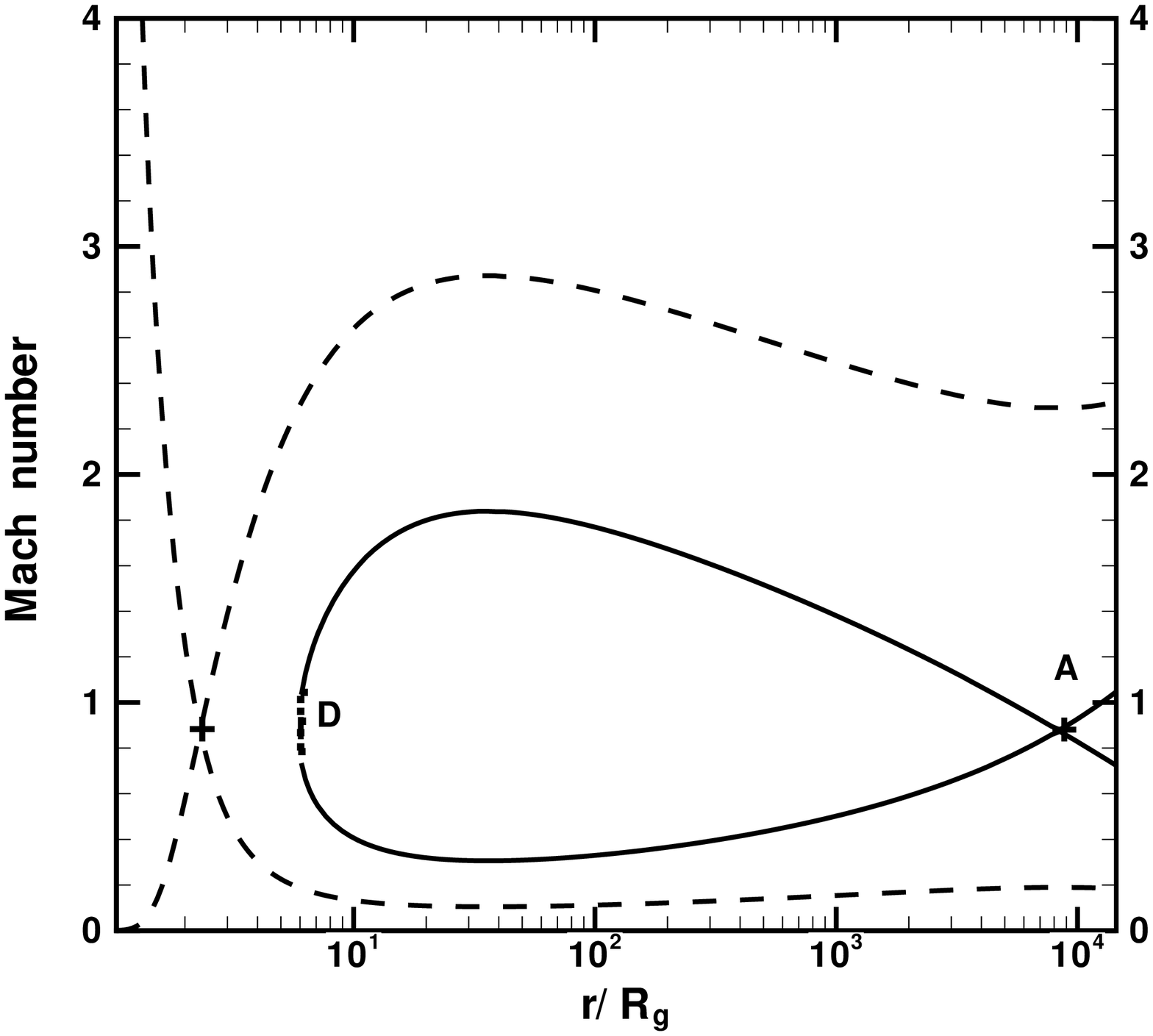}
\caption{The topology of Mach number versus radius of the transonic flow for 
  the parameters of model A, where the solid and the dashed lines intersect 
 at the outermost and innermost sonic points indicated by crosses, 
 respectively. The particle which passes through the outer sonic point A  
 falls down supersonically but never attains the event horizon.
  }
\label{fig1}
\end{minipage}
\hspace*{6pt}
\begin{minipage}{0.5\linewidth}
\includegraphics[width=7cm,height=6cm]{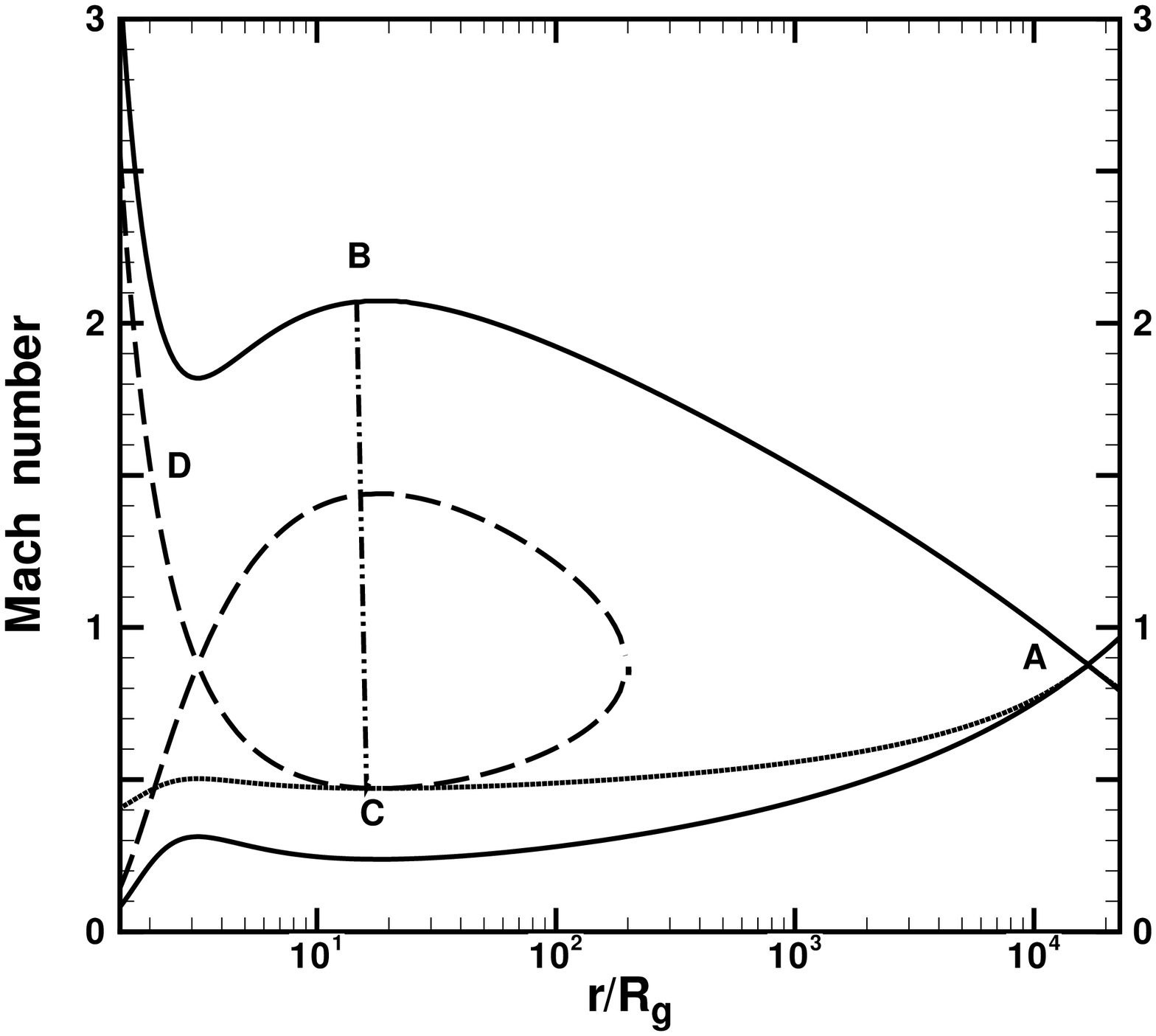}
\caption{Same as Fig.1 but for model D.
 The particle which passes through the outer sonic point A falls 
supersonically along the line AB, jumps at the shock position $R_{\rm s} \sim
20R_{\rm g}$ from the point B to the point C in a subsonic state and  again tends
towards the event horizon supersonically.  
 }
\label{fig2}
\end{minipage}
\end{figure}

\section{ Model equations}
 The set of relevant equations consists of six partial differential 
equations for density, momentum, and thermal and radiation energy.
These equations include the heating and cooling of gas and 
radiation transport.   
The radiation transport is treated in the grey, flux-limited diffusion
approximation \citep{b9}.
We use spherical polar coordinates ($r$,$\zeta$,$\varphi$), where $r$
is the radial distance, $\zeta$ is the polar angle measured from 
the equatorial plane and $\varphi$ is the azimuthal angle.  
The flow is assumed to be symmetrical with respect to the $Z$-axis 
($\partial /\partial \varphi=0 $) and the equatorial plane.
In this coordinate system, we have the basic
equations in the following conservative form \citep{b8}:

\begin{equation}
  { \partial\rho\over\partial t} + {\rm div}(\rho\bmath{v}) =  0,  
\end{equation}
\begin{equation}   
 {\partial(\rho v)\over \partial t} +{\rm div}(\rho v \bmath{v})  =
 \rho\left[{w^2\over r} + {v_\varphi^2\over r}-{GM \over (r-R_{\rm g})^2} 
 \right] -{\partial p\over \partial r}+f_r,
\end{equation}
\begin{equation}  
 {{\partial(\rho rw)}\over \partial t} +{\rm div}(\rho rw\bmath{v}) 
 = -\rho v_\varphi^2{\rm tan}\zeta-{\partial p\over\partial\zeta}
   + f_\zeta, 
\end{equation}
\begin{equation}    
{{\partial(\rho r{\rm cos}\zeta v_\varphi)}\over \partial t} 
    +{\rm div}(\rho r{\rm cos}
\zeta v_\varphi\bmath{v}) = 0, 
\end{equation}
\begin{equation}  
 {{\partial \rho\varepsilon}\over \partial t}+
   {\rm div}(\rho\varepsilon\bmath{v})
     = -p\;\rm div \bmath{v}  - \Lambda, 
\end{equation}
and
\begin{equation}       
 {{\partial E_0}\over \partial t}+ {\rm div} \bmath{F}_0+
       {\rm div}(\bmath{v}E_0 + \bmath{v}\cdot P_0) 
       = \Lambda 
     - \rho{(\kappa +\sigma)\over c}\bmath{v}\cdot
     \bmath{F}_0,
\end{equation} 
where $\rho$ is the density, $\bmath{v} =(v, w, v_\varphi)$ are the
three velocity components, $G$ is the gravitational constant,
 $p$ is the  gas pressure,
$\varepsilon$ is the specific internal energy of the gas,  $E_0$ is 
the radiation energy density per unit volume, and $P_0$ is the radiation
 stress tensor. 
We adopt the pseudo-Newtonian potential \citep{b24}
in equation (2).
The force density $\bmath{f}_{\rm R}=(f_r,f_\zeta)$ exerted 
by the radiation field is given by
\begin{equation} 
 \bmath{f}_{\rm R}=\rho\frac{\kappa+\sigma}{c}\bmath{F}_0, 
\end{equation} 
where $\kappa$ and $\sigma$ denote the absorption and scattering 
coefficients and $\bmath{F}_0$ is the radiative flux in the comoving
frame.
The quantity $\Lambda$ describes the cooling and heating of the gas, 

\begin{equation}      
     \Lambda = \rho c \kappa(S_*-E_0), 
\end{equation}
where $S_*$ is the source function.
For this source function, we assume local thermal equilibrium $S_*=aT^4$, 
where $T$ is 
the gas temperature and $a$ is the radiation constant.
For the equation of state, the gas pressure is given by the ideal gas law, 
$p=R_{\rm G}\rho T/\mu$, where $\mu$ is the mean molecular weight 
and $R_{\rm G}$ is the gas constant. 
 The temperature $T$ is proportional to the specific
internal energy, $\varepsilon$, and satisfies the relation $p=(\gamma-1)\rho\varepsilon
 =R_{\rm G}\rho T/\mu$.  
 To close the system of 
equations, we use the flux-limited diffusion approximation  
for the radiative flux:
\begin{equation}
  \bmath{F}_0= -{\lambda_0 c\over \rho(\kappa+\sigma)}
  {\rm grad}\;E_0, 
\end{equation}

\noindent and
\begin{equation}
  P_0 = E_0 \cdot T_{\rm Edd}, 
\end{equation}
where  $\lambda_0$ and $T_{\rm Edd}$ are the flux-limiter and the 
Eddington Tensor, respectively, for which we use the approximate
formula given in \citet{b8}.
The formula fulfills the correct limiting conditions $\lambda_0\rightarrow
 1/3 $ in the optically thick diffusion limit,
 and $\mid\bmath{F_0}\mid\rightarrow cE_0 $ as $\lambda_0\rightarrow 0$ 
 for the optically thin streaming limit.

In the two-temperature model with $E_0$=0 for models E and F, the energy equation (5) is replaced by the following energy equations for ion and electron.

\begin{equation}  
 {{\partial \rho\varepsilon_{\rm i}}\over \partial t}+
   {\rm div}(\rho\varepsilon_{\rm i}\bmath{v})
     = -p_{\rm i}\;\rm div \bmath{v}  - q_{\rm ie}, 
\end{equation}
and 
\begin{equation}  
 {{\partial \rho\varepsilon_{\rm e}}\over \partial t}+
   {\rm div}(\rho\varepsilon_{\rm e}\bmath{v})
     = -p_{\rm e}\;\rm div \bmath{v}   - \rho c \kappa S_* + q_{\rm ie} - q_{\rm syn}, 
\end{equation}
where $\varepsilon_{\rm i}$ and $\varepsilon_{\rm e}$ are the specific internal energy of the ion and electron, $p_{\rm i}$ and $p_{\rm e}$ are the ion and 
 electron gas pressure, $q_{\rm ie}$ is the energy 
transfer rate  from the ion to  electron by Coulomb collisions, 
$q_{\rm syn}$ is the cooling rate by synchrotron radiation \citep{b20},
 and $p$ = $p_{\rm i}$ + $p_{\rm e}$.

\section{Numerical methods}
The set of partial differential equations (1)--(6) is 
numerically solved by a finite-difference method under adequate initial 
and boundary conditions.
The numerical schemes used are basically the same as those previously described in \citet{b8} and \citet{b22}. 
These methods are based on an explicit--implicit finite difference scheme.
The inner boundary radius $R_{\rm in}$ is taken to be $2R_{\rm g}$.
In our models, the outer shock will be formed in the region
of $ r \leq 200$ $R_{\rm g}$  if it exists. 
Therefore, the outer boundary radius $R_{\rm out}$  is set to be
200$R_{\rm g}$ so that the predicted shock position never exceed the outer
boundary.
The computational domain is divided into $N_r \times N_\zeta$ grid cells,
where $N_r$ grid points (=100) in the radial direction are spaced 
logarithmically as $\Delta r/r = 0.054$ and $N_\zeta$ grid points (=100) 
in the angular direction are equally spaced.

\subsection{Initial flow variables}

The initial flow variables in the 2D time-dependent calculation for models 
A--D are given from the 1D adiabatic flow mentioned in Section 2.
To obtain the radiation energy density $E_0$, 
we specify the ratio $\beta$ of the gas pressure to the total pressure 
and the Eddington factor $f_{\rm Edd}$. 
Then, the gas temperature $T$ and the radiation energy density $E_0$ 
are given by 

\begin{equation}
 {T = \beta v_{\rm s}^2 /(R_{\rm g} \gamma)},
\end{equation}

\noindent and 

\begin{equation}
 {E_0 = (1-\beta) v_{\rm s}^2 \rho /(f_{\rm Edd} \gamma)},
\end{equation}

Concerning $\beta$, we specify its value so that the gas pressure is dominant 
and the temperature $T_{\rm B}$ at the Bondi--Hoyle--Lyttleton radius
 $R_{\rm BHL}$ ($\sim 7\times 10^4 R_{\rm g}$) is 
comparable to the assumed wind temperature $T_{\rm w}$,
where $R_{\rm BHL}$ is given by the wind velocity 1000 ${\rm km}$ s$^{-1}$
and the wind temperature $T_{\rm w}$ = 1.0 ${\rm keV}$ \citep{b17}. 
Consequently, $\beta=0.9$ and $f_{\rm Edd}=1$ are adopted in all models.
Table  2 shows the flow variables of the radial velocity 
$v_{\rm out}$, the Mach number $M_{ach}$, the density $\rho_{\rm out}$, 
the gas temperature $T_{\rm out}$, the radiation energy density 
$({E_0})_{\rm out}$, the relative thickness $h/r$ of the flow at 
the outer boundary $R_{\rm out}$, the radius  $R_{\rm sonic}$ of 
the outer sonic point and the gas temperature $T_{\rm B}$ at the 
Bondi--Hoyle--Lyttleton radius.

 On the other hand, as for the initial flow of models E and F, 
 we use the final numerical results of models A and D, respectively and  
 assume initially that $T_{\rm e}/T_{\rm i} =\delta$ and $T=(T_{\rm e} 
+ T_{\rm i})/2$ to maintain the pressure balance throughout the flow, where
 $T$ is the single temperature obtained from models A and D.

\begin{table*}
\centering
\caption{Flow variables at outer boundary $R_{\rm out}$, radius 
$R_{\rm sonic}$ of the outer sonic point, gas temperature 
$T_{\rm s}$ at the outer sonic point and gas temperature 
$T_{\rm B}$ at the Bondi--Hoyle--Lyttleton radius $R_{\rm BHL}$, 
where $\rho_0=10^{-8}$ g cm$^{-3}$. The finally obtained flow variables of
 models A and D are used as the initial flow variables of models E and F,
 respectively, but with ion temperature $(T_{\rm i})_{\rm out} =2T_{\rm out}/(1+\delta )$ and electron temperatute $(T_{\rm e})_{\rm out}=2\delta T_{\rm out}/(1+\delta)$.}
\begin{tabular}{@{}cccccccccc} \hline \hline
Model & $v_{\rm out}/c$ & $M_{ach}$ & $\rho_{\rm out}$ & $T_{\rm out}$ &
$(E_0)_{\rm out}$ & $(h/r)_{\rm out}$ & $R_{\rm sonic}$
 & $T_{\rm s}$ & $T_{\rm B}$ \\ 
 & & & (${\rm g}$ ${\rm cm}^{-3})$& ( $\rm K)$& 
(${\rho}_0 c^2)$& & ($R_{\rm g}$) & ($\rm K)$& ($\rm K$)

\\\hline
 $A$& 4.75$\times 10^{-2}$& 1.91& 5.86$\times 10^{-19}$& 2.45$\times 10^9$&
 8.86$\times 10^{-15}$    & 0.45& 8.359$\times 10^3$    & 9.5$\times 10^7$&
 9.6$\times 10^6$       \\ 
 $B$& 4.88$\times 10^{-2}$& 2.02& 5.90$\times 10^{-19}$& 2.30$\times 10^9$&
 8.35$\times 10^{-15}$    & 0.43& 1.677$\times 10^4$    & 4.7$\times 10^7$&
 1.0$\times 10^7$       \\
 $C$& 4.95$\times 10^{-2}$& 2.06& 5.83$\times 10^{-19}$& 2.28$\times 10^9$&
 8.20$\times 10^{-15}$    & 0.43& 1.681$\times 10^4$    & 4.7$\times 10^7$& 
 1.0$\times 10^7$       \\ 
 $D$& 4.97$\times 10^{-2}$& 2.07& 5.81$\times 10^{-19}$& 2.28$\times 10^9$&
 8.16$\times 10^{-15}$    & 0.45& 1.682$\times 10^4$    & 4.7$\times 10^7$&
 1.0$\times 10^7$       \\
\hline
\end{tabular}
\end{table*}

\subsection{ Initial and boundary conditions}
At the outer boundary of the accretion flow,  a continuous inflow with  
the flow variables in Table  2 is always given.
The initial atmosphere above the accreting matter in the computational domain
is  given as a radially hydrostatic equilibrium state
with zero azimuthal velocities everywhere.
Physical variables at the inner boundary, 
except for the velocities, are given by extrapolation of the variables 
near the boundary. 
 However, we impose limit conditions that the radial velocities are given 
 by a free-fall velocity and the angular velocities are zero.
 On the rotational axis and the equatorial plane, 
the meridional tangential velocity $w$ 
is zero and all scalar variables must be symmetric relative to these axes.
As for the outer boundary region above the outer accreting matter, we  use
free-floating conditions and allow for the outflow of matter, whereas here any 
inflow is prohibited.
 With these initial and boundary conditions, we perform  time 
integration of equations (1)--(6) until a quasi-steady solution is obtained. 

  \begin{figure}
  \begin{minipage}{0.5\linewidth}
      \includegraphics[width=86mm,height=66mm]{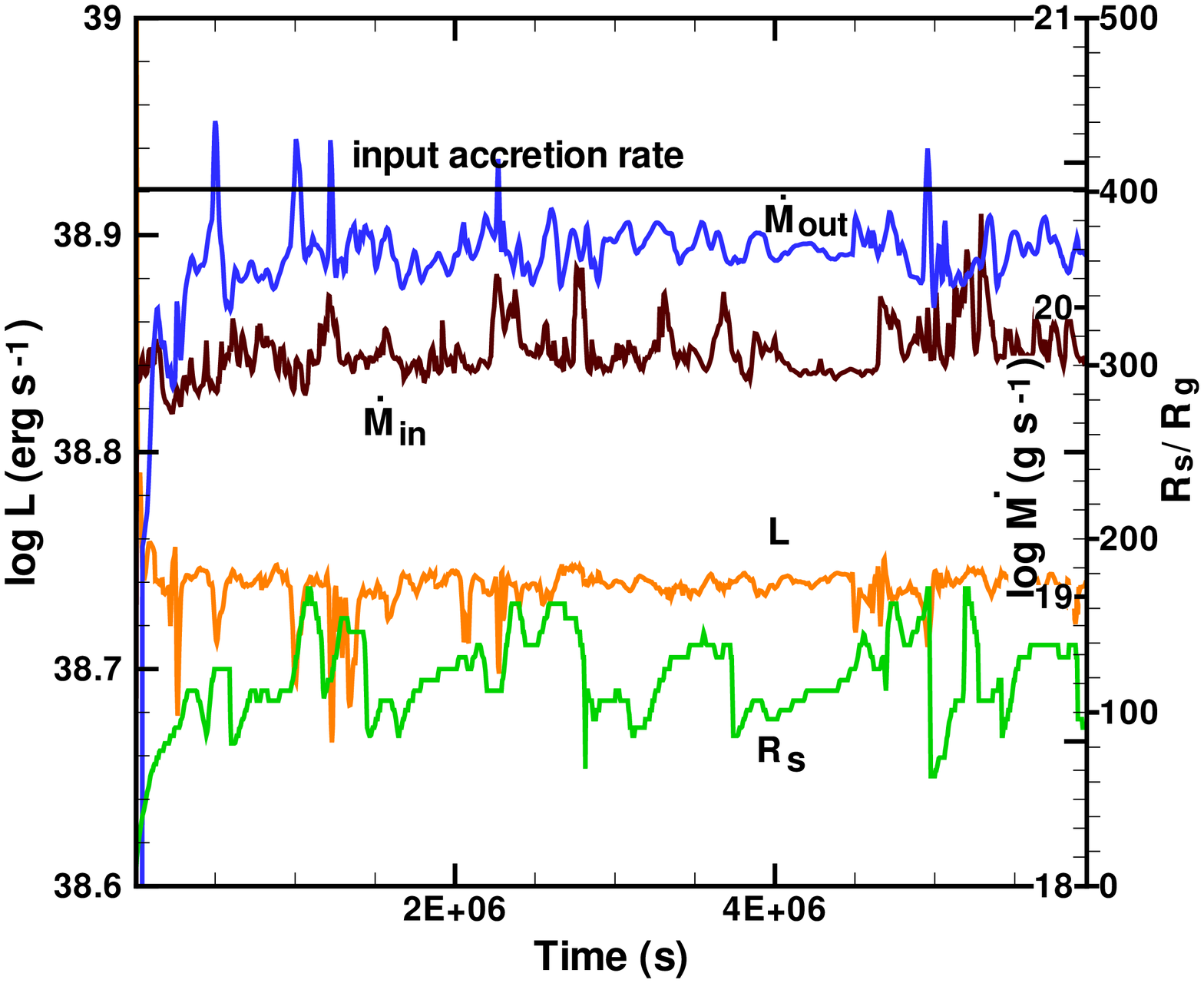}
      \caption{Time variations of total luminosity $L$,  mass-outflow rate
       $\dot M_{\rm out}$, mass-inflow rate
      $\dot M_{\rm in}$ and shock position $R_{\rm s}$ on the equatorial 
      plane for model A, where input accretion rate is indicated by a dotted 
      line and time is taken in units of seconds.}
     \label{fig3}
     \end{minipage}
  \hspace*{8pt}
  \begin{minipage}{0.5\linewidth}
     \includegraphics[width=86mm,height=66mm]{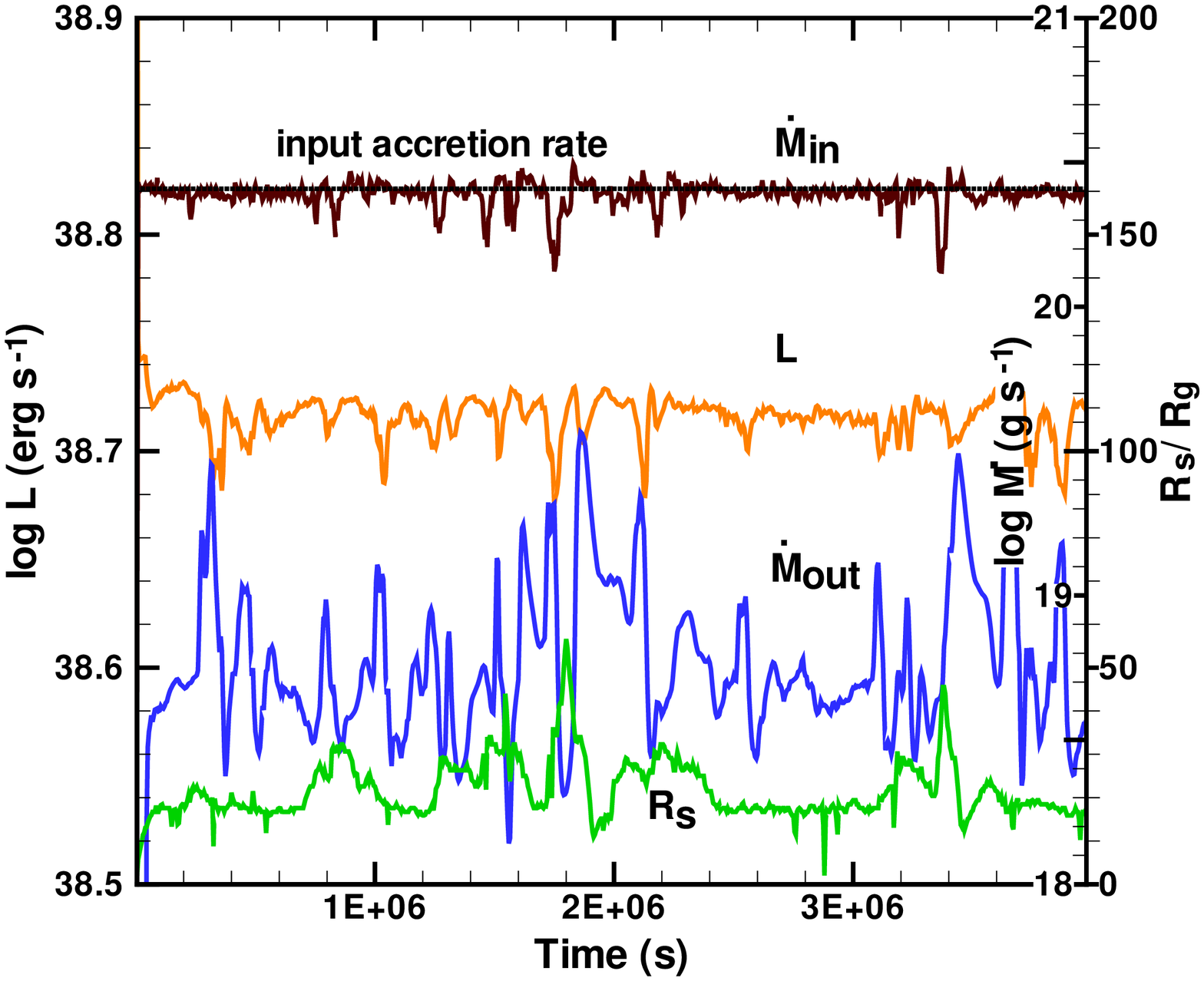}
     \caption{Same as Fig.3 but for model D.
      The luminosity is modulated by several per cent but the mass-outflow
      rate varies by a factor of 7 although it is two orders of magnitude
      smaller than the input accretion rate. }  
     \label{fig4}
 \end{minipage}
 \end{figure}

\section{Numerical results}
 To get a steady state of flow,
we need  evolutionary times longer than the following dynamical 
time $t_{\rm dyn}$.

 \begin{equation}
   t_{\rm dyn} \sim {\Omega}^{-1}  \sim {4 \times 10^4 \over \lambda}
     \left({r\over 200 R_{\rm g}}\right)^2 {R_{\rm g}\over c}, 
 \end{equation}
where $\Omega$ is the angular velocity.

This time is a measure of the needed computational time. 
Actually, because unsteady shock phenomena occurred, we needed 
more times, of 5--7 $\times 10^4 R_{\rm g}/c$, to examine the time 
variabilities, where $R_{\rm g}/c \sim 40$ s.
The total luminosity $L$ is also a good measure to check whether a steady state 
of the accretion flow is attained. The luminosity $L$ is given by 

\begin{equation}
  L = 4 \pi {R_{\rm out}}^2 \int_{0}^{\pi/2} F_r(R_{\rm out},\zeta)
      \cos{\zeta}d\zeta.
\end{equation}

\subsection{Time variations of the luminosity and the mass--flow rate}

First, we show the results of model A as a typical case.
After a time of $\sim 10^3 R_{\rm g}/c$, the total luminosity $L$ through 
the outer boundary nearly attains a steady value of $5.4 \times 10^{38}$ 
 ${\rm erg}$ ${\rm s}^{-1}$ but is modulated  only by a few to ten per cent of the
total luminosity. 
Following the time evolution of the flow, a shock wave is formed in the inner
region of $r \leq 30 R_{\rm g}$ and it moves gradually outward. 
After a time of $\sim 10^4 R_{\rm g}/c$, the shock wave begins to oscillate
irregularly between the radii of 80--160 $R_{\rm g}$  with time-scales of
hours to days.
 The mass-outflow rate $\dot
M_{\rm out}$ at the outer boundary and the mass-inflow rate $\dot M_{\rm in}$
at the inner edge are modulated  intermittently by a factor around 3.
$\dot M_{\rm out}$ plus $\dot M_{\rm in}$ is nearly equal to the input
accretion rate $\dot M_{\rm input}$ but it is remarkable that the mass-outflow
rate is twice as large as the mass-inflow rate at the inner edge,
that is, a large amount of the input accreting matter is ejected
as a wind.

Fig. 3 shows the time variations of the total luminosity $L$ (${\rm erg}$
${\rm s}^{-1}$), the mass-outflow rate $\dot M_{\rm out}$ 
(${\rm g}$ ${\rm s}^{-1}$), the mass-inflow rate $\dot M_{\rm in}$ and 
 the shock position on the equatorial plane for model A, where the input
 accretion rate  is indicated by a dotted line and time is taken in units 
 of seconds.
The shock wave moves up and down over the spatial pre-shock zone, but it
 maintains the Mach number of $\sim$ 2 throughout the shock oscillation.

\begin{table*}
\centering
\caption{ Averaged values of  $L$,  $\dot M_{\rm in}/\dot M_{\rm input}$ and 
 $\dot M_{\rm out}/\dot M_{\rm input}$, shock position $R_{\rm s}$, maximum luminosity $L_{\rm max}$ and maximum mass-outflow rate $(\dot M_{\rm out})_{\rm max}$, where $\dot M_{\rm input}$
is the input mass accretion rate.}

\begin{tabular}{@{}cccccccc} \hline \hline
Model & $L$ (${\rm erg}$ ${\rm s}^{-1}$) & $\dot M_{\rm in}/\dot M_{\rm input}$ & $\dot M_{\rm out}/\dot M_{\rm input}$ & $R_{\rm s}/R_{\rm g}$ & $L_{\max}/L$ & $ (\dot M_{\rm out})_{\rm max}/\dot M_{\rm out}$ 
 \\\hline
 $A$& 5.4$\times 10^{38}$ & 0.32            & 0.68 & 80--160  &1.04& $\sim3$ \\ 
 $B$& 5.3$\times 10^{38}$ & 1.0$\times 10^{-4}$ & 0.99 & no shock&--&-- \\
 $C$& 5.1$\times 10^{38}$ & 0.72                & 0.28 & 40--80 &1.08& $\sim 4$\\ 
 $D$& 5.1$\times10^{38}$  & 0.96                & 0.04 & 15--30 &1.05&$\sim7$\\
 $E$& 1.3$\times10^{38}$  & 0.30          & 0.70 & 70--130 &$\sim2$& $\sim2$ \\
 $F$& 1.6$\times10^{36}$  & 0.97                & 0.03 & 7--45 &$\sim5$&$\sim7$ \\
\hline
\end{tabular}
\end{table*}

Although in model B with the highest $\lambda$ the shock is formed firstly 
around $r \sim 100 R_{\rm g}$ and moves gradually upstream, it finally 
disappears and a steady flow is attained. In model C with $\lambda$ smaller 
than model A, the flow features are similar to model A. But the shock 
oscillates between 40--80 $R_{\rm g}$ and the mass-outflow rate varies by a factor of 4.
Model D differs from the other models because the initial flow 
of model D is based on the 1D adiabatic flow with stationary standing 
shocks. Fig. 4 shows the time variations of $L$,  
$\dot M_{\rm out}$, $\dot M_{\rm in}$ and $R_{\rm s}$ for model D.
First, the shock is formed  steadily at $r/R_{\rm g} \sim 20 $ 
and it begins to move between 15 $\leq r/R_{\rm g} \leq 30$.
The luminosity is modulated by several per cent but the mass-outflow
rate varies by a factor of 7 although it is two orders of magnitude
smaller than the input accretion rate. 
The Mach number of the shock maintains itself at about $\sim 4$
and is stronger than those in models A and C, because the shock is formed
closer to the black hole.

 We may wonder that the flow of model D shows non-stationary behaviours,
 because the initial flow of model D has been given from the adiabatic
 solution with the stationary standing shocks. 
 So we examined this model under an adiabatic 2D condition and 
reconfirmed that the stationary shock is formed constantly at $r 
\sim 16 R_{\rm g}$ which is slightly smaller than the theoretical values of  
$R_{\rm s}$ = 17 and 19 obtained from the 1D adiabatic solution.
In model D, unlike the adiabatic case, we notice that the cooling
and heating of the gas considerably influence the formation and 
position of the shock.

\begin{figure}
     \includegraphics[width=86mm,height=66mm]{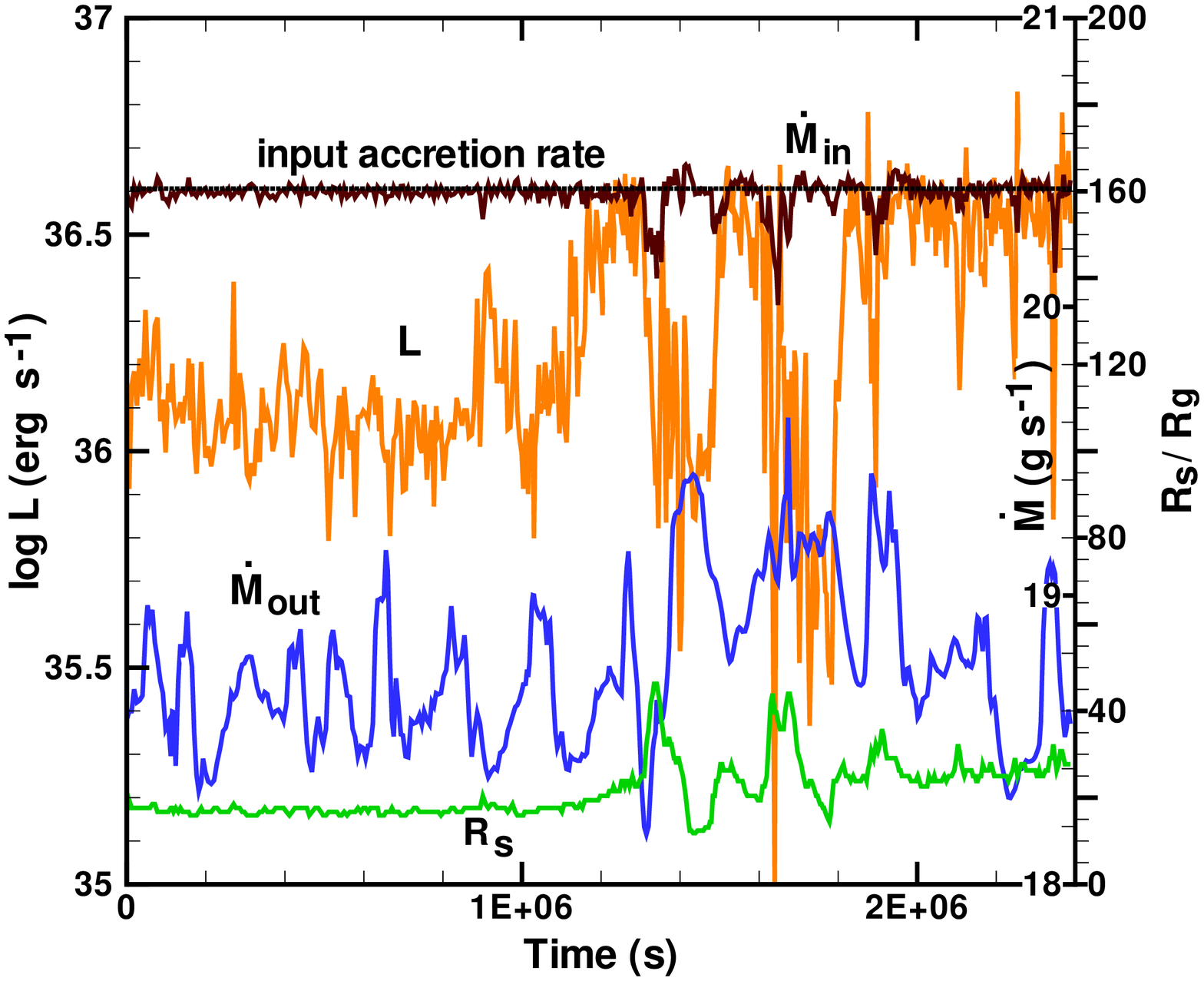}
     \caption{Time evolutions of total emission $L$ by the synchrotron
     and free-free emissions, 
       mass-outflow rate $\dot M_{\rm out}$, mass-inflow rate 
      $\dot M_{\rm in}$ and 
     shock position $R_{\rm s}$ on the equatorial plane in model F.}  
     \label{fig5}
 \end{figure}

 Fig. 5 shows the time variations of the total luminosity $L$ by
 the synchrotron and free-free emissions,  
 $\dot M_{\rm out}$, $\dot M_{\rm in}$ and $R_{\rm s}$ for model F.
 The figure also shows the irregularly oscillating shock phenomena 
 similar to model D but the modulation amplitude of the luminosity is much
 larger than that of model D and the maximum luminosity is as high as five
 times of the averaged luminosity $1.6\times 10^{36}$ erg s$^{-1}$ 
 which is comparable to the observed luminosity of Sgr A*. 
 In the two-temperature models of E and F, the synchrotron 
 emission is one orders of magnitude larger than the free-free emission and
 the energy transfer rate $q_{\rm ie}$ from the ion to electron by 
 Coulomb collisions is negligibly small compared with the synchrotron emission.

Table  3 shows the averaged values of  $L$, $\dot M_{\rm in}/\dot M_{\rm input}$ and  $\dot M_{\rm out}/\dot M_{\rm input}$, the shock position
$R_{\rm s}$ on the equatorial plane,  the maximum luminosity $L_{\rm max}$ and the maximum mass-outflow rate $(\dot M_{\rm out})_{\rm max}$, where $\dot M_{\rm input}$ is the input  accretion rate of $2.5\times 10^{20}$ g s$^{-1}$.
The luminosities of the models A--D are $\sim 5 \times 10^{38}
\;{\rm erg}\;{\rm s}^{-1}$  independently of $\lambda$.
The radiative efficiency $\eta$ $(=L/\dot M c^2)$ of the accreting matter
into the radiation is $\sim 10^{-3}$ in models A--E but is far low, 
$\sim 10^{-5}$, in model F.
The mass-outflow rate increases with increasing $\lambda$.
In model B with the highest angular momentum, most of the 
input accreting matter is ejected as the wind, without arriving at the inner 
edge, and in model D with the lowest $\lambda$, only four per cent of the 
input matter is ejected as the wind.
Conversely, the mass-inflow rate at the inner edge decreases as $\lambda$
increases.

\subsection{Structure of the flow and of the shock}

 \begin{figure}
 \begin{minipage}{0.5\linewidth}
     \includegraphics[width=86mm,height=76mm]{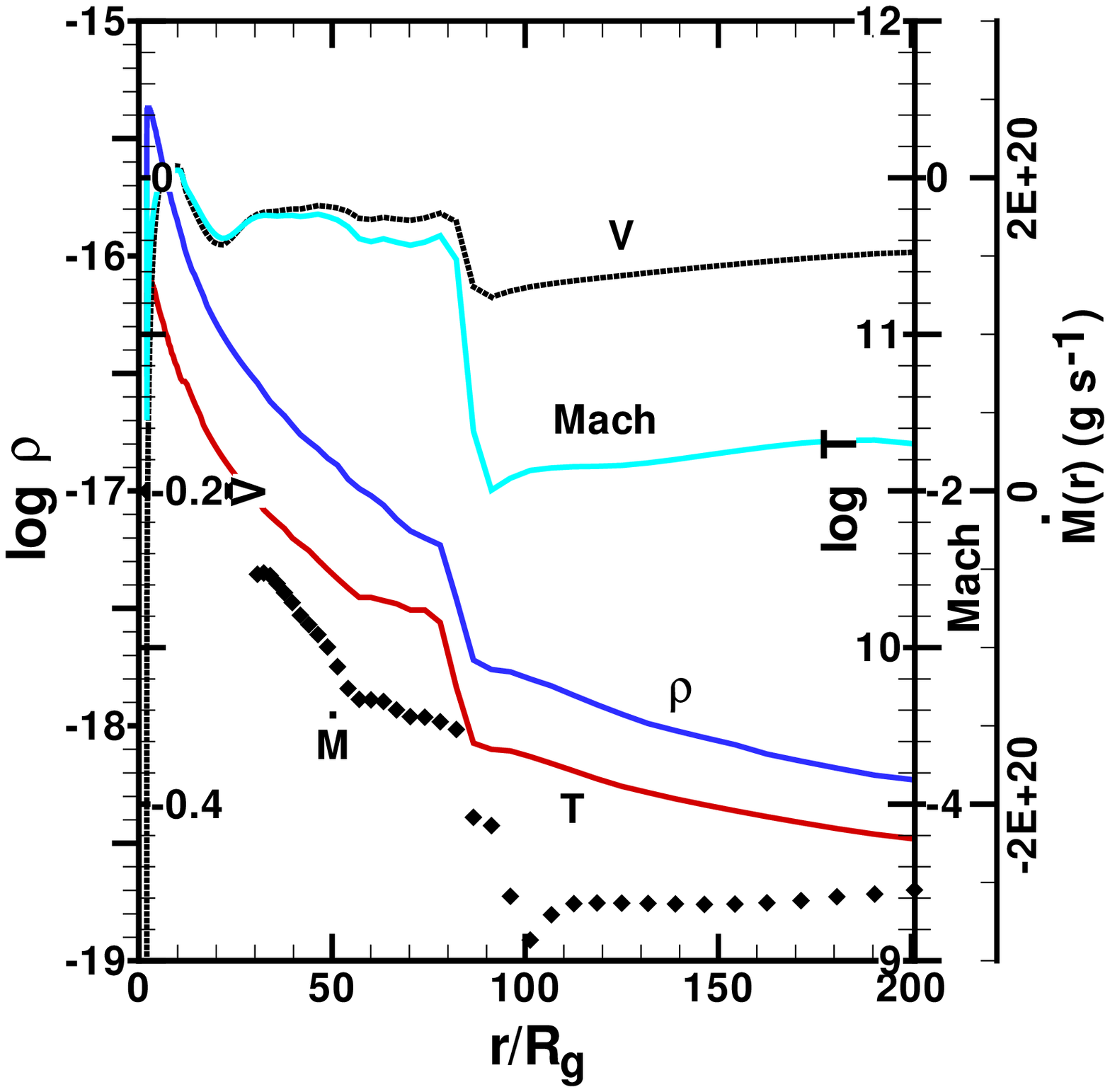}
     \caption{Profiles of density $\rho$ (g cm$^{-1}$), temperature $T$ (K),
      radial velocity $v$, Mach number of the radial velocity
      on the equatorial plane,
     and accreting mass flux $\dot M(r)$ versus radius at 
     $t = 1.7 \times 10^5$ s for model A.}  
     \label{fig6}
  \end{minipage}
  \hspace*{6pt}
 \begin{minipage}{0.5\linewidth}
     \includegraphics[width=86mm,height=76mm]{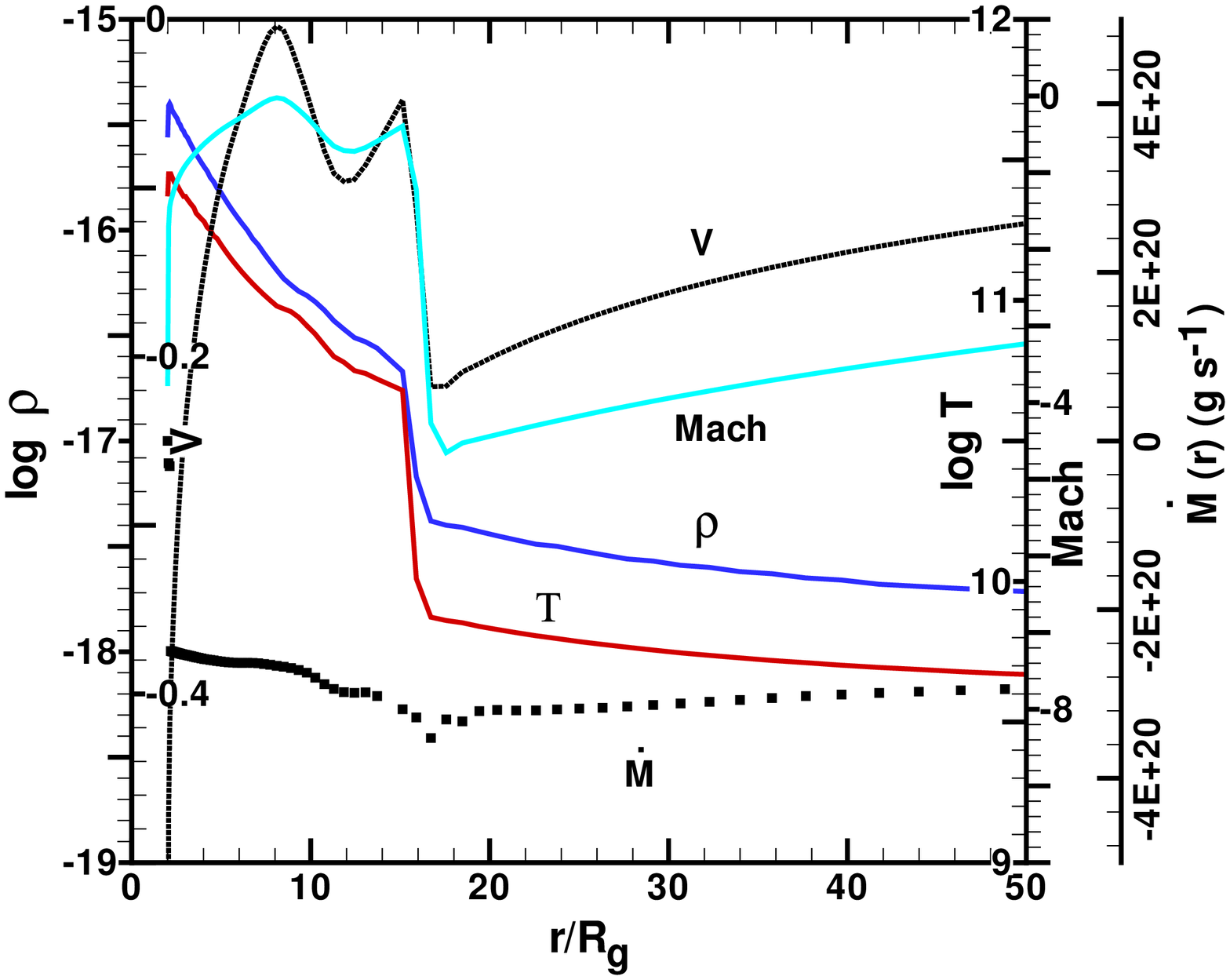}
     \caption{Same as Fig. 6 but at $t = 2.0\times 10^5$ s
     for model D. The parameters  $\lambda$ and $\varepsilon$ in this model
     correspond  to ones responsible for stationary standing shocks under 
     the adiabatic condition.}
     \label{fig7}
  \end{minipage}
  \end{figure}

Fig. 6 shows the profiles of density $\rho$ (g ${\rm cm}^{-3}$),
gas temperature $T$ (${\rm K}$), radial velocity $v$, its Mach number
on the equatorial plane and accreting mass flux $\dot M (r)$ (g s$^{-1}$) 
versus radius at $t=1.7\times 10^5$ s for model A. 
 The accreting mass flux is defined as the mass flux of the flow with
 negative radial velocities above the equatorial plane. Since the flow is 
convectively unstable in the inner region of $r \leq 30 R_{\rm g}$, $
\dot M (r)$ is not correctly defined at $r \leq 30 R_{\rm g}$ according to the above definition.
The non-stationary shock is found at a radius of 
$r \sim 90 R_{\rm g}$. The temperature $2.5 \times 10^9$ ${\rm K}$ at the outer
boundary increases to $5.0 \times 10^9$ ${\rm K}$ at the pre-shock radius
and is enhanced to $1.3 \times 10^{10}$ ${\rm K}$ across the shock. 
The density is also enhanced by a factor of three across the shock.
In Fig. 6, a single shock structure is found but we find 
two shock features at some phases.
The accreting matter falls inward with a constant accretion rate until it 
attains the shock position and then it decreases (see the scale on the 
right side of the figure) after the gas passes 
through the shock. This follows from the fact that the outflow begins in the
post-shock region after matter crosses the shock.
Fig. 7 shows the profiles of  $\rho$ (g ${\rm cm}^{-3}$),
 $T$ (${\rm K}$),  $v$, its Mach number 
and  $\dot M (r)$ (g s$^{-1}$) at $t=2.0\times 10^5$ s for model D. 
The shock on the equatorial plane is found at $r \sim 17 R_{\rm g}$. 
In this figure, we also find that the accreting mass
flux $\dot M (r)$ becomes a little small in the post-shock region although 
it is almost constant until the gas reaches the shock. This is due to the 
effect of the mass-outflow in the post-shock region. However, the effect on
the accreting mass flux is smaller than in model A, because the mass-outflow
 rate in model D is far smaller than in model A.

  \begin{figure}
     \includegraphics[width=86mm,height=76mm]{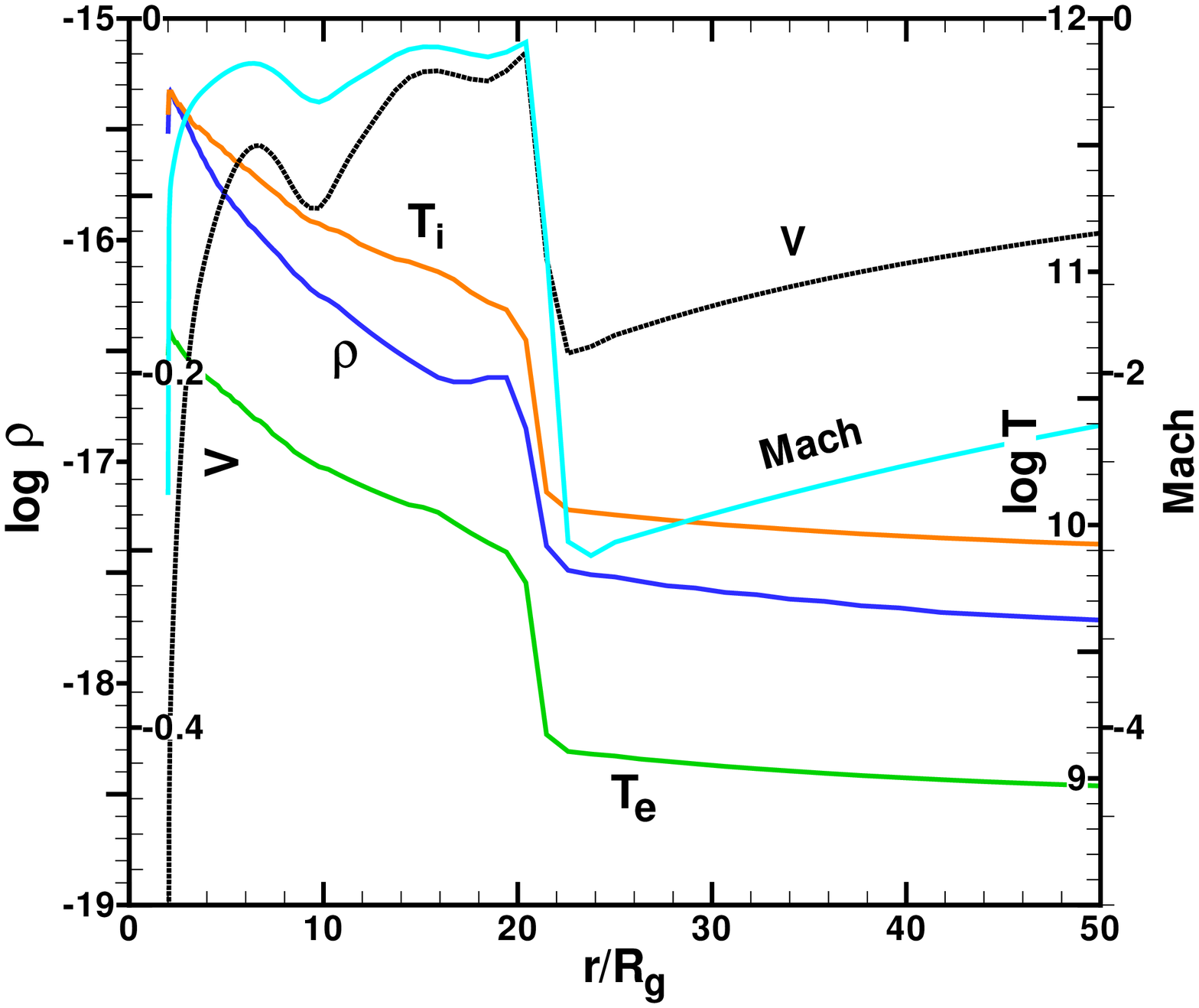}
     \caption{Profiles of density $\rho$ (g cm$^{-3}$), ion temperature 
     $T_{\rm i}$ (K), electron temperature $T_{\rm e}$ (K),
      radial velocity $v$ and Mach number of the radial velocity
      on the equatorial plane versus radius at $t = 1.3\times 10^6$ s 
     for model F.}  
     \label{fig8}
  \end{figure}

Fig. 8 shows the profiles of density $\rho$ (g ${\rm cm}^{-3}$), 
ion temperature $T_{\rm i}$ (${\rm K}$), electron temperature $T_{\rm e}$
 (${\rm K}$), radial velocity $v$,  Mach number of the velocity on the 
 equatorial plane and accreting mass flux $\dot M (r)$ (g s$^{-1}$) for 
 model F at $t=1.3\times 10^6 $ s, at which the luminosity takes a maximum
 value $3.2\times 10^{38}$ erg s$^{-1}$. The profiles in the  pre-shock region 
 are almost same as the initial one and the ratio $T_{\rm e}/T_{\rm i}$ is not
 so different from its initial ratio $\delta (=1/9)$ even in the post-shock
 region. 
 This means that the steady values of the electron and ion temperatures and 
 accordingly the intrinsic total emission of models E and F may not be still 
reached by the time integration over $ t \sim 2\times 10^6$ s.
 To obtain exact steady distribution of $T_{\rm e}$ and $T_{\rm i}$,
 the time-independent differential equations for a stationary state should be
  solved instead of the method of the time-dependent calculation.
 In spite of this fact, from the result of the two-temperature model with 
 the input accretion rate of $4.0\times 10^{-6}\;{\rm M_{\odot}}$ yr$^{-1}$,
 we can expect the total emission of $\sim 10^{36}$ erg s$^{-1}$ compatible 
 with the observed one of Sgr A*  if the electron temperatures of 
 $\sim 10^9$ -- $10^{10}$ K are obtained in the inner region.

 \begin{figure}
  \begin{minipage}{0.5\linewidth}
     \includegraphics[width=86mm,height=76mm]{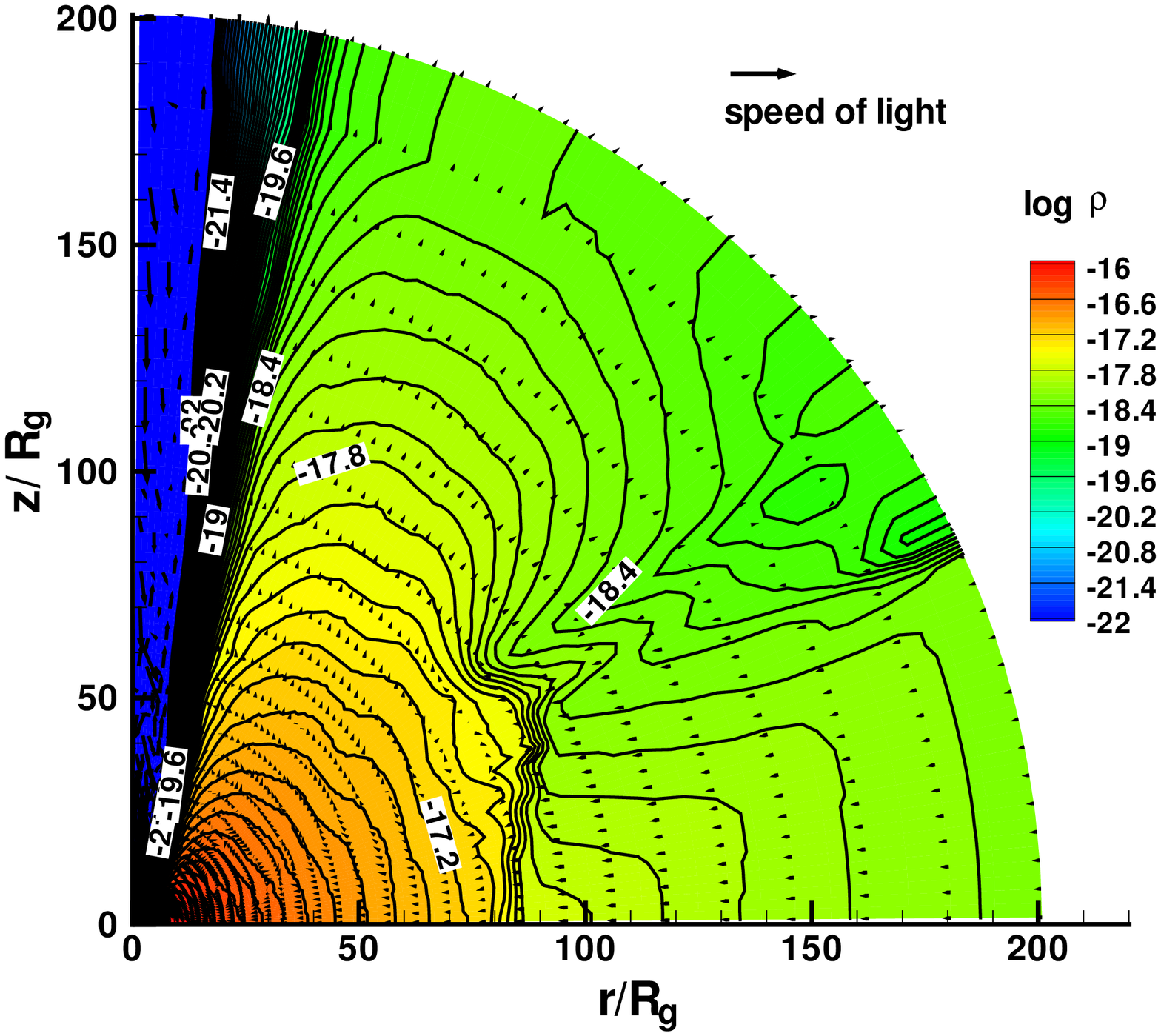}
     \caption{Contours of density $\rho$ $({\rm g}$ ${\rm cm}^{-3})$ and 
     velocity vectors at $t = 1.7\times 10^5$ s for model A. 
     The velocity of light is indicated by the upper arrow.
     The shock wave extending obliquely is formed at $r/R_{\rm g}\sim$ 85 
     up to a height of $z \sim 50 R_{\rm g}$.
     }  
     \label{fig9}
  \end{minipage}
  \hspace*{6pt}
  \begin{minipage}{0.5\linewidth}
     \includegraphics[width=86mm,height=76mm]{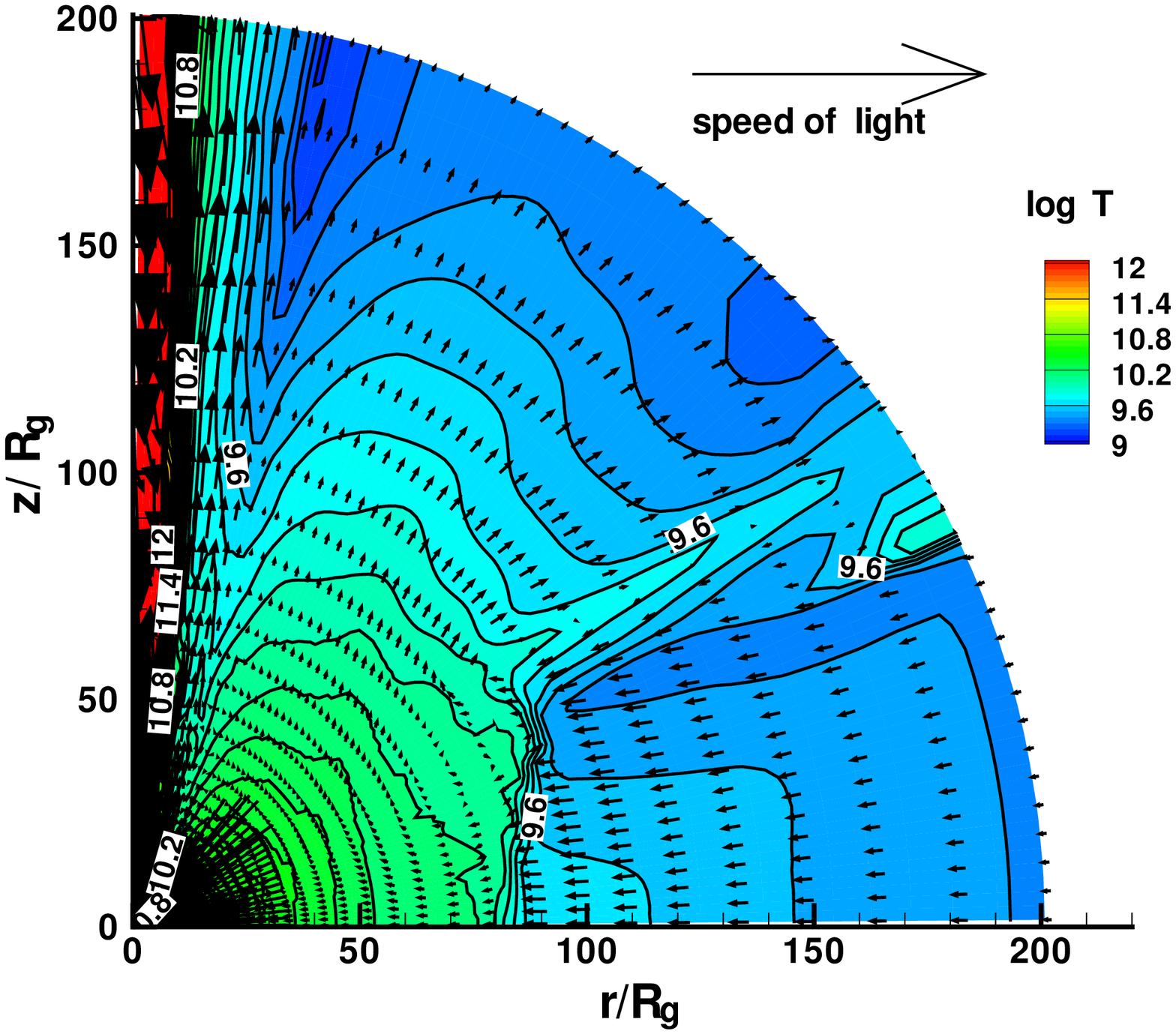}
     \caption{Same as Fig. 9 except for the temperature, where the velocity
     vectors are amplified by five times, compared with Fig. 9. }  
     \label{fig10}
     \end{minipage}
   \end{figure}

 Figs 9 and 10 show the contours of density $\rho$ 
 (g cm$^{-3}$) and temperature $T$ (K) with velocity vectors at 
 $t=1.7 \times 10^5$ s in model A.
The shock wave rises obliquely at $r \sim 90 R_{\rm g}$ 
above the equatorial plane and extends upward to $\sim 50 R_{\rm g}$.
The velocity vectors in Fig. 10 are amplified by five times, compared with 
Fig. 9.   The initial accreting matter at the outer boundary supersonically
falls down towards the gravitating centre and is decelerated at the shock 
front, enhancing the density and the temperature and becomes subsonic.
The post-shock region with high densities and high temperatures 
results in a strong outward pressure-gradient force also along the 
$Z$-axis and a part of the accreting matter deviates from the disc flow 
and escapes as the wind flow. 
The wind velocity becomes $0.01$--$0.05c$ at the outer boundary except for a 
narrow funnel region along the $Z$-axis, where the funnel wall is 
characterized by the vanishing of the effective potential and the flow velocity
 within the funnel region is relativistic at $\sim 0.2c$.
The remaining accreting matter is swallowed into the 
central black hole. In model A, two thirds of the input accreting matter 
is expelled from the accretion flow.
The mass-ouflow begins after the matter undergoes the shock compression. 
The circulating flow is also found at $r \leq 30 R_{\rm g}$, 
that is, the accretion flow is convectively unstable in the post-shock region.

\subsection{Properties of the time-variability  of the flow}
The most remarkable feature of the low angular momentum flow considered
here is the existence of the non-stationary and oscillating shocks found 
in all models except for model B. This shock phenomenon results in the irregular modulation of the flow variable.
.

\begin{figure}
\begin{minipage}{0.5\linewidth}
\includegraphics[width=66mm,height=66mm]{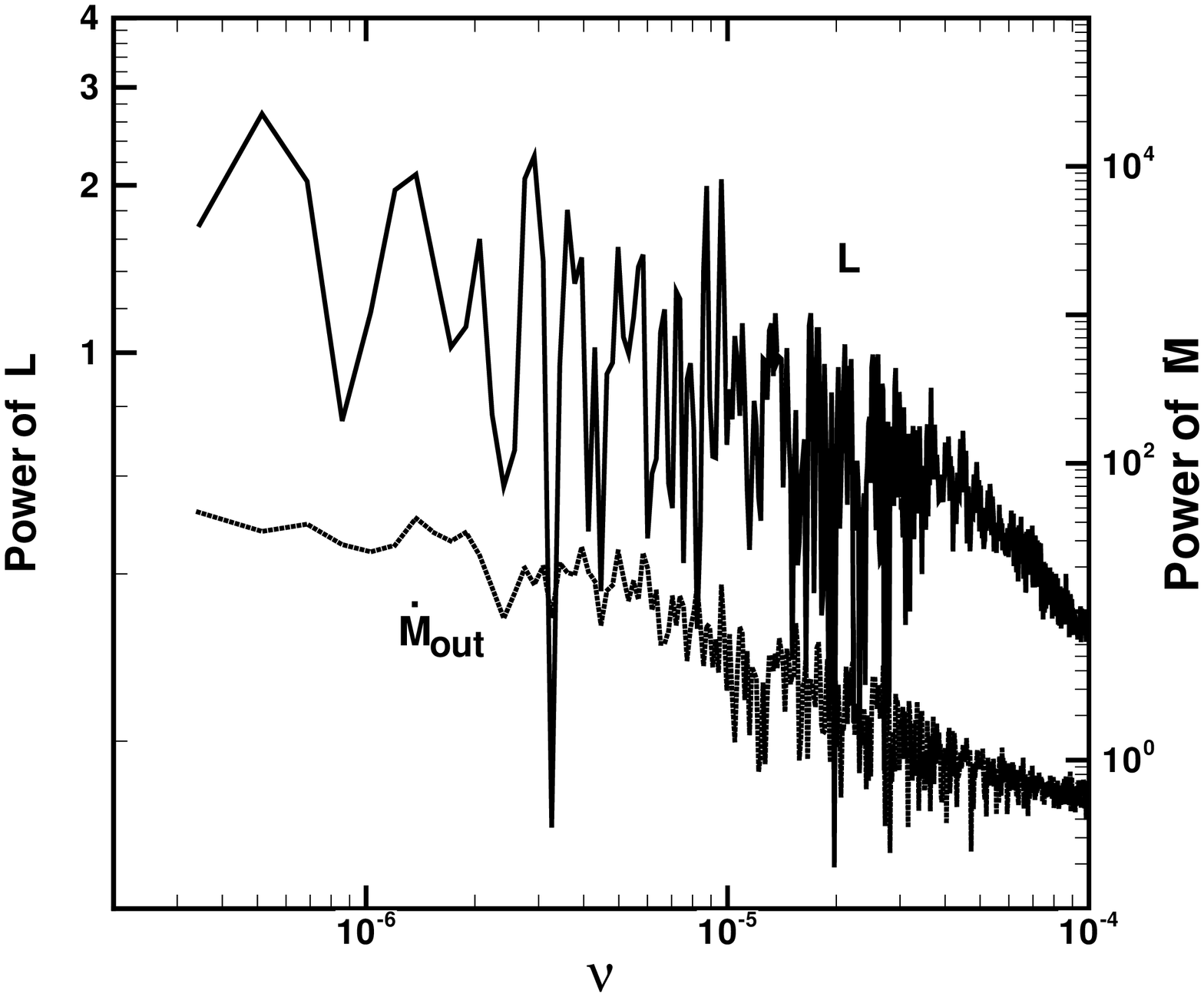}
\caption{Power spectrum density of the total luminosity $L$ 
      and the mass-outflow rate $\dot M_{\rm out}$ in model A.
     }  
\label{fig11}
\end{minipage}
\hspace*{6pt}
\begin{minipage}{0.5\linewidth}
\includegraphics[width=66mm,height=66mm]{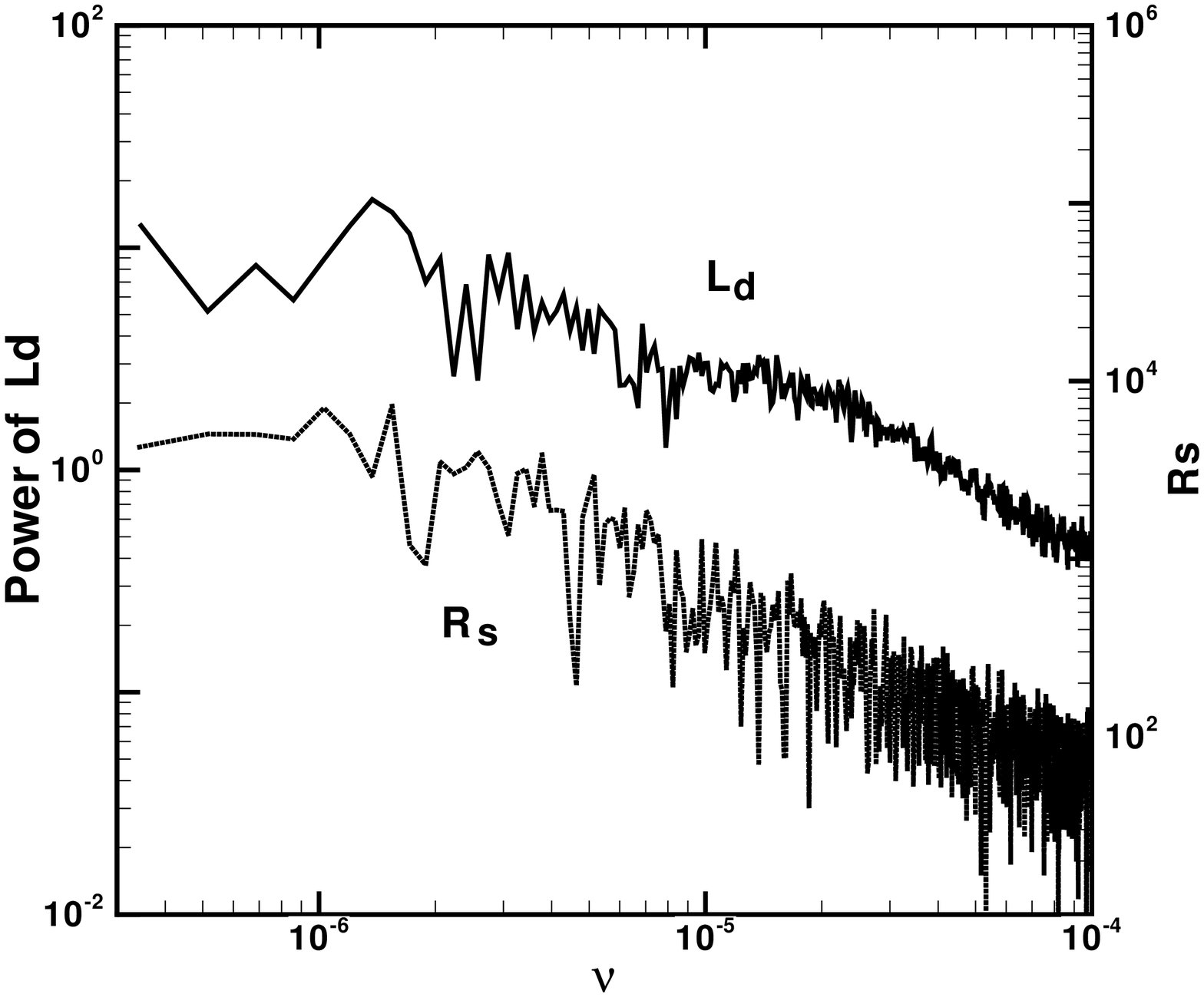}
\caption{Same as Fig. 11 but for disc luminosity $L_{\rm d}$ 
      and the shock position  $R_{\rm s}$ on the equatorial plane.
     }  
\label{fig12}
\end{minipage}
\end{figure}

The luminosity varies synchronously with the mass-outflow rate
without any time lag. 
In spite of the large modulation of the shock position in models A, C, 
and D, the modulation amplitude of the luminosity is as small as a few 
to ten per cent , while in models E and F it varies by  a factor of  2--5. 
The outflow consists of a persistent outflow and a large episodic one with 
irregular intervals of an hour to ten days and the episodic outflow rate
is higher by a factor of 2--7 than the averaged mass-outflow rate.

Sgr A* possesses a quiescent state and a flare state. 
The flares of Sgr A* have been detected in multiple wavebands from radio, 
sub-millimetre, IR to X-ray \citep{b33,b34}. The amplitudes of the variability at radio, IR and X-ray are by factors of below 1/2, 1--5 and 45 or even
higher, respectively, a few times per day. 
The rapid variability in the radio band on time-scales of a few seconds to 
hours has been observed \citep{b34}.  
Although a quasi-periodicity of $\sim$ 20 minutes in the IR flares 
\citep{b5,b12,b13,b28,b7} and radio QPOs of 16, 22, 31 and 56 minutes 
\citep{b15} have been reported, they are the subject of a controversy \citep{b14,b4}.
These flare phenomena are interesting in relation to the time variations of
the luminosity and the mass-outflow rate in the present models. 
We consider that the intermittently large modulations of the outflow due
to the oscillatory shock may be related to the flares of Sgr A*.
Unfortunately, the time resolution in our calculations is at most as small 
as a few minutes because of the limited time step in the calculations. 
Therefore, the short time variability at a scale of  $\sim $ 10 minutes
 observed in Sgr A* can not be compared with the result of our models.
 Figs 11 and 12 show the power spectrum density of the total luminosity $L$ 
and  mass-outflow rate $\dot M_{\rm out}$ and of the disc luminosity 
$L_{\rm d}$ and  shock radius $R_{\rm s}$ on the equatorial plane, 
respectively, for model A, where the disc luminosity is defined at the surface 
of the inward flow with only negative radial velocities near the equatorial
 plane.
Here, we can not find a conspicuous frequency peak in these power spectra 
 partly due to insufficient statistical power but
they are suggestive of semi-regular variabilities on time-scales 
of hours to days.

The origin of the oscillating shock appearing in the present models is 
entirely attributed to the range of parameters of the specific energy and 
the specific angular momentum used here. 
Generally, the shock formation and its position are very sensitive to
the upstream flow parameters.
The initial flow variables of the models, except for models D and F with 
$\lambda=1.35$, are based on the 1D adiabatic flow in which parameters  
$\lambda$ never produce a stationary standing shock but are located just
 nearby in the range of parameters responsible for the stationary shock.
Furthermore, we know that the shock  becomes unstable when we take account of
 the energy loss and gain of the gas \citep{b16,b23}.

\section{Comparison with the advection-dominated accretion flow}
The present low angular momentum flow models are compared
 with the ADAF model \citep{b19,b20,b21,b31} which is referred to as
 a radiatively inefficient accretion flow model of Sgr A. 
The region considered here is limited to an inner region of $r \leq 200R_{\rm g}$.
The ADAF has generally the following  properties: (1) the accretion
flow becomes geometrically thick, (2) the pressure support 
in the radial direction is considerable, 
so the angular velocity is sub-Keplerian, (3) the radial velocity of the gas
is much larger than that of the standard disc model and is barely less than
Keplerian velocity in the innermost region, and (4) the large radial velocity
and the large scale height cause the gas density to be very low, and so 
the cooling time is very long compared with the dynamical time and 
the matter is optically thin. 
These properties are also realized in the present low angular momentum flows.
The ADAF solution is gas-pressure dominated and the sound velocity is 
comparable to the Keplerian velocity, so the gas temperature becomes nearly
virial. 
Gas at such a high temperature radiates copiously at small radii where
the temperature can approach $10^{12} {\rm K}$. Thus, in the ADAF model, 
a two-temperature plasma of the ion temperature $T_{\rm i}$ and the electron
temperature $T_{\rm e}$  is assumed to exist at the small radii.
Typical ADAF models have the two temperatures scaling as \citep{b21}
\begin{equation}
 T_{\rm i} \sim 10^{12}/r \;\;{\rm K}, 
\;\;\; T_{\rm e} \sim {\rm Min}(\;T_{\rm i},10^{9-11}\;\;{\rm K}) .
\end{equation}
In our models A--D, the two-temperature model is not considered but a high 
temperature of $\sim 2 \times 10^{11}$ K comparable to the above ion
temperature is obtained in the innermost region.
The temperature profile is roughly $T \propto r^{-1}$ in all models,
in agreement with the ADAF. 
 In the ADAF with the constant accretion rate, a self-similar flow with the
 density profile of $\rho (r) \propto r^{-s} $
 gives $s=1.5$, while our models show $s=1.4$--$1.8$.
The radiative efficiency of the accretion flow 
in our model F is as small as $\sim 10^{-5}$ in the usual ADAFs.

Thus, as far as the inner region of the accretion flow is concerned,
the low angular momentum flow model has many properties in common with those of the ADAFs.
The distinction between the ADAFs and the present low angular momentum flow 
 is that the former describes a hot, nearly spherical, accretion flow
 with viscosity, while the latter shows the same accretion flow with lower
angular momentum but without viscosity.
The remarkable difference between our models and the ADAFs is
the existence of the oscillating shock phenomena in the low angular momentum
 model, except for model B with a relatively larger angular momentum. 
The shock position $R_{\rm s}$ is irregularly oscillatory in the inner 
region of $ r \leq 180 R_{\rm g}$. 
Another distinctive point of our models is that the mass-outflow is 
considerable, depending on the specific angular momentum of the model,
as mentioned in Section 5.1.

\begin{figure}
     \includegraphics[width=86mm,height=66mm]{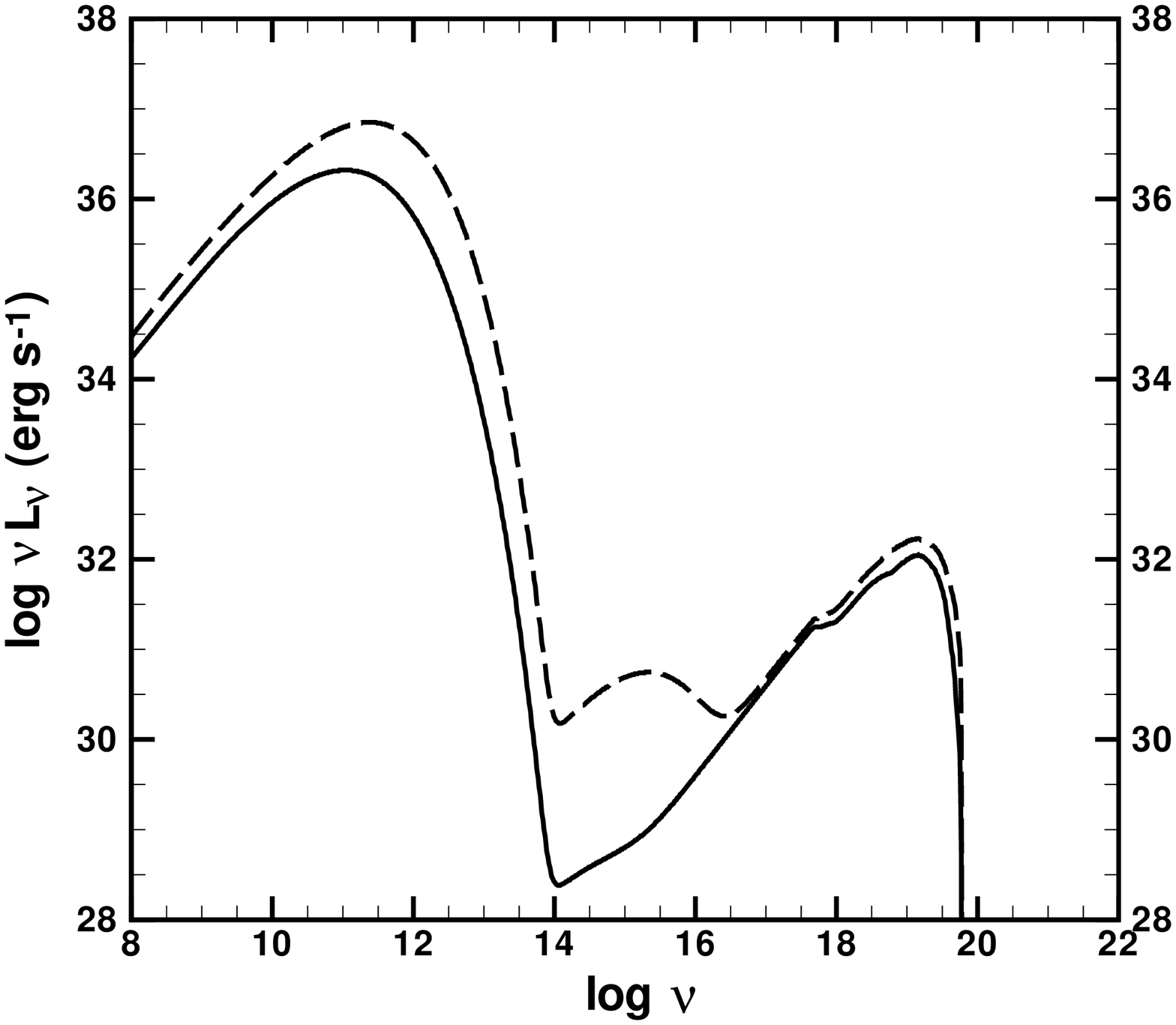}
     \caption{Energy spectra $\nu L_{\rm \nu}$ ${\rm erg}$
      ${\rm s}^{-1}$ of model F  with $ L=3.2\times
     10^{36}$ erg s$^{-1}$ at $t=1.3\times 10^6$ s (dashed line)
       and with 
      $L=1.2\times 10^{36}$ erg s$^{-1}$ $t=7.1\times 10^5$ s 
      (solid line)}.   
     \label{fig13}
 \end{figure}

 Fig. 13 shows the energy spectra calculated from model F 
 with $ L=3.2\times 10^{36}$ erg s$^{-1}$ at $t=1.3\times 10^6$ s (dashed line) and with $L=1.2\times 10^{36}$ erg s$^{-1}$ at $t=7.1\times 10^5$ s 
(solid line).   
 The sub-milimetre and X-ray bumps in these spectra are originated from 
 the synchrotron cooling by thermal electrons and the bremsstrahlung cooling,
 respectively, whose formulas are prescribed in \citet{b20} and \citet{b27}.
 The intensities in the sub-milimetre bumps of models E and F are a little
 higher and the peak positions lie in a longer wavelength compared with the 
 observational spectra at the quiescent state of Sgr A*.
 To the contrary, the intensities in the X-ray band (2--8 keV) are one 
 orders of magnitude lower than the observed one.
 The radiatively inefficient accretion flow models \citep{b31},
 such as the ADAFs, take into account the complicated physics including 
 synchrotron and inverse Compton emission by thermal and non-thermal electrons
 under the two-temperature assumption and describe the energy
 spectra very fitted to the observed ones of Sgr A* over a wide range of energy
 bands from the radio to X-ray. 
 In this respect, the present our model is too simple and insufficient to
  reproduce the spectra of Sgr A* but it is  useful to examine the basic 
 scenario of the low angular momentum flow model.

\section{Summary and discussion}
Based on the analysis of the angular momentum of the accretion flow around 
Sgr A* by \citet{b17}, we examined a low angular momentum flow model for
Sgr A* using two-dimensional  hydrodynamical calculations.
The models with the specific angular momenta $\lambda$ of 1.35, 1.55, 1.68, 
and 2.16 (in the usual nondimensional units) and input accretion rates of 
$\dot M_{\rm input} = 4.0\times 10^{-6}$ was considered.  We summarize the results as follows.

(1) The accretion flow is very hot, optically thin, geometrically thick,
   and convectively unstable in the inner region of the accretion flow,
   as is also found in the ADAFs.

(2) The flow is highly advected and, if the syncrotron cooling 
and a two-temperature model are taken into account, the radiative efficiency 
of the accreting matter into radiation is very low, $\sim 10^{-5}$--$10^{-3}$,
and the input accretion rate
$\dot M_{\rm input} = 4.0\times 10^{-6}\;{\rm M_{\odot}}$ yr$^{-1}$ results in 
 the total luminosity $\sim 10^{36}$  ${\rm erg}$ ${\rm s}^{-1}$ which is 
 comparable to the observed  luminosity of Sgr A*.

(3) An irregularly oscillating shock is formed in the inner region of a few
  tens to a hundred and sixty Schwarzschild radii, except for the model with
 the highest $\lambda = 2.16$.
 Due to the oscillating shock, the luminosity and  the mass-outflow
rate are modulated by several per cent to a factor of  5 and a factor of  2-7, 
 respectively, on time-scales of an hour to ten days.

(4) The mass-outflow rate increases with increasing $\lambda$ and it ranges
over a few per cent, one-thirds, two-thirds and 99 per cent, of the input
 accreting matter, corresponding to the cases of $\lambda$ = 1.35, 1.55, 1.68
 and 2.16, respectively. 

(5) The oscillaing shock is necessarily triggered if the specific angular momentum $\lambda$ and the specific energy $\varepsilon$ belong to or are located just nearby in the range of parameters responsible for a stationary  shock in the rotating inviscid and  adiabatic accretion flow.

(6) The time variability may be related to the flare phenomena of Sgr A*.

Since the specific angular momenta used in the models are 
reasonably estimated, the time variability of Sgr A* may be attributed
 naturally to the oscillating shock in the low angular momentum model
considered here. Further evidence of the low angular momentum of the accretion
 flow around Sgr A* would confirm the validity of the models.
 If the specific angular mommentum of the accretion flow around Sgr A* is
 actually as small as $\lambda=1.35$, the time variability itself of the 
 total emission like model F may explain the quiescent and the flare states of
 Sgr A*. 
 On the other hand, in the case of higher angular momentum i.e. $\lambda=1.68$,
 the time modulation of the total emission $L$ is  small to explain 
 the flare phenomena of Sgr A*.
 However, under this circumstance , we expect that the large episodic 
 wind caused by the non-stationary shock may induce the flare phenomena
 of Sgr A*. 
 The flare phenomena of Sgr A* have been seen to be polarized in the radio, 
 sub-millimetre and IR bands \citep{b5,b12,b13,b28,b6}. 
 This is a strong evidence for a magnetic field around Sgr A*. 
 The episodic wind in our model may be accelerated as a relativistic jet 
 by some mechanism such as magnetic reconnection process in a current 
 seet under the magnetic field away from the accretion flow \citep{b30}. 
 The jet will collide with the ambient matter and excite the matter as a hot 
 emitter, especially at times of the episodic  wind.
 The emission from the hot matter may be observed as the flares in X-ray.
 Furthermore, the jet expands and cools in the far distant interstellar 
 matter and results in the synchrotron emission due to 
 the electrons in the magnetic field. 
 This may be observed as the radio flares with a time lag from the X-ray flares. The existence of the magnetic field plays an important role in the accretion
 flow. However, the magnetic field was not taken into account simultaneously
 with the hydrodynamics in our models.
 Thus, the current low angular momentum flow model is simple and can not 
reproduce the observed spectra of Sgr A*, as is successfully fitted over a
 wide range of the wavelength by ADAFs.
 For a spectral fitting, we have to treat the physics of the synchrotron
 and inverse Compton emission by the thermal and non-thermal electrons under
 the two-temperature model. 
 A further magneto-hydrodynamical examination of the flow with a low 
angular momentum model including the above physics should be pursued 
 in the future.

\label{lastpage}

\end{document}